\documentstyle[prd,aps,epsf]{revtex} 

\begin{document}

\draft
\title{		Scalar-Composite Model in $6-2\varepsilon$ Dimensions}
\author{			Keiichi Akama$^1$ and Takashi Hattori$^2$}
\address{	$^1$Department of Physics, Saitama Medical University,
 			 Saitama, 350-0496, Japan}
\address{	$^2$Department of Physics, Kanagawa Dental College,
                     Kanagawa, 238-8580, Japan}
\date{\today}
\maketitle

\begin{abstract}
We study the model of a composite-scalar made of a pair of scalar fields 
	in $6-2\varepsilon$ dimensions, using equivalence to the renormalizable 
	three-elementary-scalar model under the ``compositeness condition."
In this model, the composite-scalar field is induced by the quantum effects through the 
	vacuum polarization of elementary-scalar fields with $2N$ species.
We first investigate scale dependences of the coupling constant and masses,
	in the renormalizable three-elementary-scalar model,
	and derive the results for the composite model 
	by imposing the compositeness condition.
The model exhibits the formerly found general property 
	that the coupling constant of the composite field is independent of the scale. 
\end{abstract}


\section{Introduction}

The composite fields induced by quantum fluctuations 
	have wide application in physics,
	in connection with or without spontaneously broken symmetries.
These ideas were used in Ginzburg-Landau theory and BCS model 
	of superconductivity \cite{BCS} and 
	Nambu-Jona-Lasinio type model \cite{NJL} of hadron physics, and
	were also applied to gauge bosons \cite{ig}, Higgs scalar, 
	quarks, leptons \cite{comp}, induced gravity \cite{iG}, 
	braneworld \cite{BW}, nuclear physics, solid state physics, cosmology, etc..
Many of these composite models are not renormalizable, and
	the non-renormalizable nature plays essential roles 
	in establishing compositeness and further properties of the composites. 
Though they are not renormalizable, 
	we can formulate them through corresponding renormalizable field theory, 
	and then transfer to the composite model 
	by imposing compositeness condition $Z_3=0$ \cite{CC}, where $Z_3$ 
	is the renormalization constant of the to-be-composite field.
So far, many interesting studies have been performed on this line \cite{history}.
In the previous paper \cite{ScaleComp}, 
	we showed scale independence of the effective coupling constants 
	of the composite fields,
	in three examples of models: 
	(i) scalar composite in six dimensions, 
	(ii) Nambu-Jona-Lasinio model in four dimensions, and 
	(iii) induced gauge theory in four dimensions.
This property of scale independence 
	seems to common to the many of the quantum composite models,
	and exhibits sharp contrast with the scale dependence 
	of the running coupling constants in the elementary field theories.

In this paper, we discuss further details of 
	the scalar-composite model in six dimensions.
This model resembles the Nambu-Jona-Lasinio model for fermions 
	in four dimensions,
	serving as a toy model without fermionic complexities.
The scalar fields in six dimensions also attract attentions
	in connection with the braneworld theories in extra dimensions.
For example, they are expected to provide models for solitonic confinement 
	of the fields within our four dimensional spacetime.

The plan for this paper is as follows.
In Sec. II, we define the system to clarify, 
	what we mean by the "compositeness condition".
In Sec. III, we present the formulation in the leading order 
	in $1/N$ for illustrations.
The detailed formulation in the next-to-leading order is given in Sec.\ IV,
	and Sec.\ V is devoted to conclusions and discussions.

\section{Scalar-composite model in $6-2\varepsilon$ dimensions}

Strictly speaking, this is a scalar-composite-scalar model, which means a model of  
	a composite scalar field made of two elementary scalar fields.
We can well investigate this model by using equivalence to 
	the renormalizable three-elementary-scalar model under 
	the compositeness condition.

\subsection{Scalar-composite-scalar model}

We consider a composite complex scalar field $\Phi$ comprised 
	of elementary complex scalar fields $ X^1$ and $ X^2$ 
	in $6-2\varepsilon$ dimensions, where $\varepsilon$ 
	is a small positive number.
We suppose that the fields $X^i \ (i=1,2)$ have $N$-plicated components
\begin{eqnarray}
	X^i=(X_1^i, \cdots, X_N^i),
\end{eqnarray}
	for we use $1/N$ expansion latter.
We assume that the basic Lagrangian ${\cal L}$ of model is given by
\begin{eqnarray}
	{\cal L}= \sum_{i=1}^2 ( |\partial_\mu X^i |^2-m_X^2 |X^i |^2 )
	-F | X^1 X^2 |^2,
\label{CompL}
\end{eqnarray}
	where $m_X$ is the mass of $X^i$, and $F$ is the coupling constant.
Here and hereafter, summations over $N$-plicated components 
	are implied in each pair of $X^i$.
The Lagrangian (\ref{CompL}) is invariant under the transformations of 
	SU$ (N) \otimes$U$_1(1) \otimes$U$_2(1) \otimes$Z$_2$,
	where SU$(N)$ is the group simultaneously (for $X^1$ and $X^2$) 
	operating on the space of $N$ components of $X^i$,
	U$_i$(1) ($i$=1 and 2) are the groups separately (for $X^1$ and $X^2$)
	generated by the number operators of $X^i$,
	and Z$_2$ is the discrete symmetry group 
	of exchange of the fields $X^1$ and $X^2$.
Because of the four-scalar interaction $|X^1 X^2|^2$ with positive mass dimension, 
	the model is non-renormalizable.
However the summation over the chain diagrams in Fig.~1 
	give rise to a pole in the momentum squared of the channel 
	as far as we fix $\varepsilon$ at a small but non-vanishing value.
It corresponds to a composite state of the elementary fields $X^1$ and $X^2$.
To deal with the composite state, we introduce an auxiliary field $\Phi$.
Then the Lagrangian (\ref{CompL}) is equivalent to 
\begin{eqnarray}
	{\cal L}' = \sum_{i=1}^2 ( |\partial_\mu X^i |^2-m_X^2 |X^i |^2 )
	-F^{-1}|\Phi|^2 + \Phi (X^1 X^2) + {\rm h.c. .}
\label{AuxL}
\end{eqnarray}
The equivalence between the Lagrangians (\ref{CompL})
	 and (\ref{AuxL}) is shown by coincidence of 
	the generating functionals for Green functions 
	derived from (\ref{CompL}) and (\ref{AuxL}).

This model resembles the Nambu-Jona-Lasinio model \cite{NJL} in many aspects, 
	though the fermions in the latter is replaced by the scalars $X^i$ here, 
	and the number of space-time dimensions are different.
They have four-field interactions with a coupling constant of mass-dimension $-2$.
They have a composite states arising from the poles of the chain diagram, 
	and their coupling constants are dimensionless.
The Nambu-Jona-Lasinio model can be well analyzed by using its equivalence to 
	the Yukawa-type renormalizable model under the compositeness condition \cite{CC}.
Similarly we can use the equivalence of the present model to the 
	three-elementary-scalar model, as will be demonstrated in the next section.

\subsection{Renormalizable three-elementary-scalar model}

Let us consider the following renormalizable model in $6-2\varepsilon$ dimensions:
\begin{eqnarray}
	\tilde{\cal L} 
	= \sum_{i=1}^2 ( |\partial_\mu \chi_0^i |^2-m_0^2 |\chi_0^i |^2 ) 
	+ |\partial_\mu \phi_0 |^2
	-M_0^2 |\phi_0|^2 + g_0 \phi_0 \chi_0^1 \chi_0^2+{\rm h.c. ,}
\label{BareL}
\end{eqnarray}
	where $\phi_0$ and $\chi_0^i \ (i=1,2)$ 
	are elementary complex scalar fields, 
	$g_0$ is a coupling constant, and 
	$m_0$ and $M_0$ are the masses of $\chi_0^i$ and $\phi_0$ respectively.
The fields $\chi_0^i$ have $N$-plicated components
\begin{eqnarray}
	\chi_0^i=(\chi_{01}^i, \cdots, \chi_{0N}^i).
\end{eqnarray}
The subscript "$0$" denotes the bare quantities distinguished 
	from renormalized ones which will appear latter.
The Lagrangian (\ref{BareL}) does not include terms quartic in $\phi_0$ and $\chi_0$ 
	and higher, because they have positive mass dimensions 
	in the $6-2\varepsilon$ dimensions, and violates renormalizability.
The Lagrangian (\ref{BareL}) is the most general renormalizable form 
	with the above mentioned symmetry
	SU$ (N) \otimes$U$_1(1) \otimes$U$_2(1) \otimes$Z$_2$
	of the original model (\ref{CompL}).

Now let us define renormalized quantities
\begin{eqnarray}
	g &=& Z_1^{-1} Z_2 Z_3^{{1 \over 2}} g_0 \mu^{-\varepsilon},
\label{ZrenG}
\\
	\chi^i &=& Z_2^{-{1 \over 2}} \chi_0^i,
\label{ZrenChi}
\\
	\phi &=& Z_3^{-{1 \over 2}} \phi_0,
\label{ZrenPhi}
\\
	m &=& Z_m^{-{1 \over 2}} Z_2^{{1 \over 2}} m_0,
\label{Zrenm}
\\
	M &=& Z_M^{-{1 \over 2}} Z_3^{{1 \over 2}} M_0,
\label{ZrenM}
\end{eqnarray}
	for the bare quantities $g_0$, $\chi_0^i$, $\phi_0$, 
	$m_0$ and $M_0$, respectively, 
	where $Z_1$, $Z_2$, $Z_3$, $Z_m$ and $Z_M$ 
	are the renormalization constants.
In (\ref{ZrenG}), the scale parameter $\mu$ with mass dimension 
	is introduced to make the renormalized coupling constant $g$ dimensionless.
Then the Lagrangian (\ref{BareL}) is rewritten as
\begin{eqnarray}
	\tilde{\cal L} 
	= \sum_{i=1}^2 ( Z_2 |\partial_\mu \chi^i |^2 -Z_m m^2 |\chi^i |^2 )
	+ Z_3 |\partial_\mu \phi |^2 - Z_M M^2 |\phi|^2 
	+ Z_1 g \mu^\varepsilon \phi \chi^1 \chi^2+{\rm h.c..}
\label{RenL}
\end{eqnarray}
We define the renormalized Lagrangian $\tilde{\cal L}_{\rm r}$ by
\begin{eqnarray}
	\tilde{\cal L}_{\rm r} 
	&=& \sum_{i=1}^2 \left(|\partial_\mu \chi^i|^2-m^2|\chi^i|^2 \right)
	+|\partial_\mu \phi|^2-M^2|\phi|^2
	+g \mu^\varepsilon \phi \chi^1 \chi^2 + {\rm h.c.,}
\end{eqnarray}
and the counter Lagrangian $\tilde{\cal L}_{\rm c}$ by
\begin{eqnarray}
	\tilde{\cal L}_{\rm c} 
	&=& \sum_{i=1}^2 
	\left[(Z_2-1)|\partial_\mu \chi^i|^2-(Z_m-1)m^2|\chi^i|^2 \right]
	+(Z_3-1)|\partial_\mu \phi|^2-(Z_M-1)M^2|\phi|^2
\cr
&&
	+(Z_1-1)g \mu^\varepsilon \phi \chi^1 \chi^2 + {\rm h.c. ,}
\end{eqnarray}
	so that
\begin{eqnarray}
	\tilde{\cal L}= \tilde{\cal L}_{\rm r}+\tilde{\cal L}_{\rm c}.
\label{RenLL}
\end{eqnarray}
Thus we can calculate physical quantities in this system by perturbative method.

Let us consider the scale dependence of this model.
The renormalization group equation is given by 
\begin{eqnarray}
	\left[ \mu {\partial \over \partial \mu}
	+\beta_g {\partial \over \partial g}
	+\beta_m {\partial \over \partial m}
	+\beta_M {\partial \over \partial M}
	-n_\chi \gamma_\chi -n_\phi \gamma_\phi \right] {\tilde \Gamma}^{(n)}
	 (p_1, 	\cdots, p_n, g, m, M, \mu) =0,
\label{RenGroupEq}
\end{eqnarray}
	where ${\tilde \Gamma}^{(n)} (p_1, \cdots, p_n, g, m, M, \mu)$ 
	is a one-particle-irreducible Green function 
	with a total of $n(=n_\chi+n_\phi)$ external lines, 
	where $n_\chi$ and $n_\phi$ are the numbers of 
$\chi^i$ and $\phi$ in ${\tilde \Gamma}^{(n)}$, respectively.
In (\ref{RenGroupEq}), 
	the functions $\beta_g$, $\beta_m$, $\beta_M$ 
	are defined by
\begin{eqnarray}
	\beta_g &=& \mu {\partial g \over \partial \mu},
\label{betaG}
\\	\beta_m &=& \mu {\partial m \over \partial \mu},
\label{betam}
\\	\beta_M &=& \mu {\partial M \over \partial \mu},
\label{betaM}
\end{eqnarray}
	where functions $\beta_g$, $\beta_m$ and $\beta_M$ 
	give the scale dependence of the coupling constant $g$, 
	the masses $m$ and $M$, respectively.
In (\ref{RenGroupEq}), $\gamma_\chi$ and $\gamma_\phi$ 
	are anomalous dimensions to the fields 
	$\chi$ and $\phi$, respectively, 
	and are defined by
\begin{eqnarray}
	\gamma_\chi &=& {1 \over 2}\mu {\partial \over \partial \mu} \ln Z_2,
\label{gammaChi}
\\	\gamma_\phi &=& {1 \over 2}\mu {\partial \over \partial \mu} \ln Z_3,
\label{gammaPhi}
\end{eqnarray}
	where $Z_2$ and $Z_3$ are the renormalization constants.

\subsection{Compositeness condition}

Let us compare the scalar-composite-scalar model in the subsection A
	and the renormalizable three-elementary-scalar model in the subsection B.
The Lagrangian (\ref{RenL}) of the latter coincides with Lagrangian (\ref{AuxL}) 
	of the former, if the condition
\begin{eqnarray}
	Z_3=0, \ \ \ Z_1 \neq 0, \ \ \ 
	Z_2 \neq 0, \ \ \ 
	Z_m \neq 0, \ \ \ 
	Z_M \neq 0
\label{CC}
\end{eqnarray}
holds, and if we identify 
\begin{eqnarray}
	\chi^i &{\rm \ with\ }& {X_i \over \sqrt{Z_2}},
\label{id_chi}\\
	m^2  &{\rm \ with\ }& {Z_2 \over Z_m}m_X^2,
\label{id_m2}\\
	\phi &{\rm \ with\ }& {Z_2 \over Z_1}\left({ \Phi \over g \mu^\varepsilon} \right), {\rm \ and\ }
\label{id_phi}\\
	M^2  &{\rm \ with\ }& {Z_1^2 \over Z_2^2 Z_M}\left({ g^2 \mu^{2\varepsilon} \over F} \right).
\label{id_M2}
\end{eqnarray}
Therefore, we can calculate all the physical quantities 
	in the system with (\ref{AuxL}) at non-vanishing $\varepsilon$ 
	via Lagrangian (\ref{RenLL}) of the three-elementary-scalar model 
	with the condition (\ref{CC}).
This description includes the physical field $\phi$ 
	which does not exist in the original elements of the system with (\ref{CompL}).
This means that the field $\phi$ is a composite in the system.
Under the condition (\ref{CC}), the system (\ref{RenLL}) describes the case 
	where the field $\phi$ is a composite.
Hence the condition (\ref{CC}) is called ``compositeness condition'' \cite{CC}, 
	which gives a constraint on the parameters of the renormalizable 
	three-elementary-scalar model (\ref{BareL}).
According to (\ref{ZrenG})-(\ref{ZrenM}), the relations (\ref{id_chi})-(\ref{id_M2}) 
	can be transformed into those for the bare quantities,
\begin{eqnarray}
	\chi_0^i  &=&  X_i,
\label{chi0}
\\
	m_0^2  &=&  m_X^2,
\label{m02}
\\
	\phi_0  &=&  {\Phi \over g_0},
\label{phi0}
\\
	M_0^2  &=&  {g_0^2 \over F}.
\label{M02}
\end{eqnarray}
As for compositeness condition (\ref{CC}), the relation (\ref{ZrenG}) indicates 
	that it corresponds to the limiting case
\begin{eqnarray}
	g_0 \rightarrow \infty, 
\label{ComLimt}
\end{eqnarray}
as far as $Z_1$, $Z_2$ and $g$ are finite and non-vanishing,
	as is the case of our present interest.
We can further see from (\ref{M02})
\begin{eqnarray}
	M_0 \rightarrow \infty, 
\label{ComLimtM0}
\end{eqnarray}
for finite $F$.
Thus the compositeness condition  (\ref{CC}) is realized only 
	in the limiting case of the bare parameters.
It is, however, not at all a drawback because it is sufficient 
	that the physical renormalized quantities are finite, as they are in fact.

\section{Leading order in $1/N$}

Now we examine the scale dependences of the coupling constant $g$ 
	and the masses $m$ and $M$ in the model.
In this chapter we present the formulation in the leading order in $1/N$ 
	for illustration, where we assign $g^2 \sim \varepsilon N^{-1}$.
We first consider those in the three-elementary-scalar model in subsection A and 
	then we impose the compositeness condition to obtain the results for 
	the scalar-composite-scalar model in subsection B.

\subsection{Renormalization of the three-elementary-scalar model}

We first calculate the renormalization constants in leading order in $1/N$, 
	and we investigate the renormalization group flow 
	of the three-elementary-scalar model (\ref{BareL}).
In $1/N$ expansion, the leading order contributions come 
	from the self-energy part of $\phi$ 
	with one $\chi^i$-loop in Fig.~2 because of $g^2 \propto 1/N $.
The renormalization constants are calculated as
\begin{eqnarray}
	Z_1  &=&  Z_2\ =\ 1,
\label{Zlead12}
\\	Z_3  &=&  1-{Ng^2 \over 6(4\pi)^3 \varepsilon},
\label{Zlead3}
\\	Z_m  &=&  1,
\label{Zleadm}
\\	Z_M  &=&  1-{Ng^2 m^2 \over 6(4\pi)^3 \varepsilon M^2},
\label{ZleadM}
\end{eqnarray}
	in minimal subtraction scheme.
Hereafter we adopt the minimal subtraction scheme throughout this paper.

Now we discuss the renormalization group flow of the 
	three-elementary-scalar model.
>From the renormalization constant (\ref{Zlead3}), 
	we obtain the function $\beta_g$ in the leading order,
\begin{eqnarray}
	\beta_g = -\varepsilon g 
	\left[1-{Ng^2 \over 6(4\pi)^3 \varepsilon} \right].
\label{betaGlead}
\end{eqnarray}
According to (\ref{betaG}) and (\ref{betaGlead}), 
	we have the differential equation,
\begin{eqnarray}
	\mu {\partial g \over \partial \mu}
	= -\varepsilon g 
	\left[1-{Ng^2 \over 6(4\pi)^3 \varepsilon} \right].
\label{DeffLeadG}
\end{eqnarray}
It is solved to give the scale dependence of coupling constant $g$,
\begin{eqnarray}
	g^2 = \left[ {\mu^{2\varepsilon} \over g_0^2}
	+{N \over 6(4\pi)^3 \varepsilon} \right]^{-1},
\label{GdependLead}
\end{eqnarray}
	where we have chosen $1/g_0^2$ as the integration constant 
	in accordance with (\ref{ZrenG}).
Then (\ref{GdependLead}) gives the renormalization group flow 
	of the coupling constant $g$ in the three-elementary-scalar model 
	at the leading order.
Substituting (\ref{Zlead12}) and (\ref{Zlead3}) 
	to (\ref{gammaChi}) and (\ref{gammaPhi}), 
	the anomalous dimensions $\gamma_\chi$ and $\gamma_\phi$ are given by 
\begin{eqnarray}
	\gamma_\chi   &=&  0,
\label{gammaChiS}
\\	\gamma_\phi   &=&   {Ng^2 \over 6(4\pi)^3}.
\label{gammaPhiS}
\end{eqnarray}

Next, we consider the scale dependence of the masses $m$ and $M$.
According to (\ref{Zlead12}), (\ref{Zleadm}), (\ref{ZleadM}), (\ref{gammaChiS}) 
	and (\ref{gammaPhiS}), we obtain the functions $\beta_m$ and $\beta_M$,
\begin{eqnarray}
	\beta_m  &=&  0,
\label{betamlead}
\\	\beta_M  &=&  {Ng^2 \over 6(4\pi)^3 M}(M^2-6m^2),
\label{betaMlead}
\end{eqnarray}
	where $g^2$ is given by (\ref{GdependLead}).
Using (\ref{betamlead}) and (\ref{betaMlead}) for (\ref{betam}) and (\ref{betaM}), 
	we have the differential equations as 
\begin{eqnarray}
	\mu {\partial m \over \partial \mu}  &=&  0,
\label{DeffLeadm}
\\	\mu {\partial M \over \partial \mu} 
	  &=&   {Ng^2 \over 6(4\pi)^3 M}(M^2-6m^2).
\label{DeffLeadM}
\end{eqnarray}
>From (\ref{DeffLeadm}), the mass $m$ is a constant in this order:
\begin{eqnarray}
	m^2=m_0^2,
\end{eqnarray}
	where $m_0$ is the bare mass of $\chi_0^i$ in (\ref{BareL}).
The eq.~(\ref{DeffLeadM}) is solved to give the scale dependence of mass $M$,
\begin{eqnarray}
	M^2 = {Ng^2 \over (4\pi)^3 \varepsilon}m_0^2 
	+ \left[1- {Ng^2 \over 6(4\pi)^3 \varepsilon} \right] M_0^2,
\label{MDependLead}
\end{eqnarray}
	where $g^2$ is given by (\ref{GdependLead}), 
	and we have chosen $M_0^2$ as the integration constant 
	in accordance with (\ref{ZrenM}).
The relation (\ref{MDependLead}) gives the renormalization group flow 
	for the mass $M$ in the leading order.

\subsection{The composite model in the leading order}

Here we see that the coupling constant $g$ is independent of the scale $\mu$ 
	in the scalar-composite-scalar model \cite{ScaleComp}.
If we use the compositeness condition (\ref{CC}) 
	with (\ref{Zlead3}), we have
\begin{eqnarray}
	g^2 = {6(4\pi)^3 \varepsilon \over N}.
\label{CompG}
\end{eqnarray}
We can directly see that the form of $g^2$ in (\ref{CompG}) 
	is independent of the scale $\mu$.
Substituting (\ref{CompG}) into (\ref{betaGlead}), 
	we find
\begin{eqnarray}
	\beta_g =0,
\label{betaComp}
\end{eqnarray}
	which implies that the scale independence 
	is realized as a smooth limit of the running coupling constant 
	in general non-composite case.
In fact we can see that the scale dependent parts 
	in the solution (\ref{GdependLead}) smoothly disappear 
	in the compositeness limit (\ref{ComLimt}).

At first sight, it looks like that $M$ is independent of the scale
	according to (\ref{MDependLead}) with (\ref{CompG}). 
This is, however, not the case 
	because (\ref{GdependLead}) indicates that	 
	the factors in the second term of $M^2$ in (\ref{MDependLead}) are 
\begin{eqnarray}&&
	1- {Ng^2 \over 6(4\pi)^3 \varepsilon}
	={6(4\pi)^3 \varepsilon \mu^{2\varepsilon}\over N{g_0}^2}
	+O\left(1\over g_0{}^{4}\right) \rightarrow 0
\label{1stfactor}
\end{eqnarray}
	in the compositeness limit (\ref{ComLimt}) and (\ref{ComLimtM0}), 
	and therefore the second term in (\ref{MDependLead}) has indefinite form 
	in the limit.
Using (\ref{M02}), we finally obtain the expression 
\begin{eqnarray}
	M^2 = 6{m_0}^2 
	+ \frac{6(4\pi)^3\varepsilon\mu^{2\varepsilon}}{NF},
\label{M2mudep}
\end{eqnarray}
	which is finite and non-vanishing. 
Thus we can see that $M^2$ still depends on the scale parameter $\mu$
	even in the compositeness limit.

It would be interesting to see what happens if we impose the condition 
\begin{eqnarray}
	Z_M=0
\label{ZM=0}
\end{eqnarray}
	in addition to the compositeness condition $Z_3=0$. 
The identification (\ref{id_M2}) indicates that 
\begin{eqnarray}
	F \rightarrow \infty, 
\label{F=infty}
\end{eqnarray}
	and (\ref{M2mudep}) reduces to  
\begin{eqnarray}
	M^2 = 6{m}^2. 
\label{M2=6m2}
\end{eqnarray}
Hence the condition (\ref{ZM=0}) imposes the definite relation (\ref{M2=6m2})
	among the renormalized masses $M$ and $m$,
	and makes the composite mass $M$ independent of the scale $\mu$.

\section{Next-to-Leading Order in $1/N$}

Now we present detailed formulations at the next-to-leading order in $1/N$.

\subsection{Renormalization of the three-elementary-scalar-model}

The next-to-leading order contributions to the renormalization parts 
	come from the diagrams in Fig.~\ref{f3} and Fig.~\ref{f4}.
The blobs on the $\phi$-propagators indicate 
	arbitrary numbers of $\chi$-loop insertions as shown in Fig.~\ref{f5}.
We have to take into account the multi-loop insertions 
	because  we assign $g^2 \sim O(N^{-1})$.
In Appendix A, we show the calculations of the leading divergence of each diagram
	in the renormalization parts.
Using (\ref{Sdiv}), we write the renormalization conditions 
	for the self-energy part of $\chi^i$ 
	(in the minimal subtraction scheme, as is adopted in this paper) as 
\begin{eqnarray}&&
	{1 \over N}\sum_{n=0}^\infty \left[{Ng^2 \over 6(4\pi)^3 \varepsilon} \right]^n 
	+Z_2-1 =0,
\label{Z2Ren}
\\&&	{3 \over N}\sum_{n=0}^\infty \left[{Ng^2 \over 6(4\pi)^3 \varepsilon} \right]^n
	\left[nM^2-(6n-7)m^2 \right] + (Z_m-1)m^2 =0.
\label{ZmRen}
\end{eqnarray}
Then we obtain the renormalization constants $Z_2$ and $Z_m$ from (\ref{Z2Ren}) 
	and (\ref{ZmRen}), respectively : 
\begin{eqnarray}
	Z_2   &=&   1+{1 \over N}\ln \left[1-{Ng^2 \over 6(4\pi)^3 \varepsilon} \right],
\label{ZnextChi}
\\	Z_m   &=&   1 + {3g^2 \over (4\pi)^3 \varepsilon} \left(1- {M^2 \over 6m^2} \right)
	\left[1- {Ng^2 \over 6(4\pi)^3 \varepsilon} \right]^{-1}
	+ {21 \over N} \ln \left[1-{Ng^2 \over 6(4\pi)^3 \varepsilon } \right].
\label{Znextm}
\end{eqnarray}
Using (\ref{Pidiv}), we write the renormalization conditions 
	for the self-energy part of $\phi$ as
\begin{eqnarray}&&
	{2 \over N}\sum_{n=1}^\infty \left[{Ng^2 \over 6(4\pi)^3 \varepsilon} \right]^{n+1}
	\left({1 \over n}-{1 \over n+1} \right)+ Z_3 -1 =0,
\label{Z3Ren}
\\&&	{2 \over N}\sum_{n=1}^\infty \left[{Ng^2 \over 6(4\pi)^3 \varepsilon} \right]^{n+1}
	\left[{1 \over n+1}(108m^2-9M^2)-{54 \over n}m^2 \right] + (Z_M-1)M^2 =0.
\label{ZMRen}
\end{eqnarray}
Then we obtain the renormalization constants $Z_3$ and $Z_M$ from (\ref{Z3Ren}) 
	and (\ref{ZMRen}), respectively :
\begin{eqnarray}
	Z_3  &=&  1- {g^2 \over 6(4\pi)^3 \varepsilon}(N+2)
	- {2 \over N} \left[ 1-{Ng^2 \over 6(4\pi)^3 \varepsilon} \right]
	\ln \left[1-{Ng^2 \over 6(4\pi)^3 \varepsilon} \right],
\label{ZnextPhi}
\\
	Z_M  &=&  1- {Ng^2 m^2 \over (4\pi)^3 \varepsilon M^2}
	- {3g^2 \over (4\pi)^3 \varepsilon} \left(1-{12m^2 \over M^2} \right)
\cr&&	- {18 \over N} 
	\left[1-{12m^2 \over M^2}+{Ng^2 m^2 \over (4\pi)^3 \varepsilon M^2} \right]
	\ln \left[1-{Ng^2 \over 6(4\pi)^3 \varepsilon} \right]. \ \ 
\label{ZnextM}
\end{eqnarray}
In this model, we have no vertex correction at this order, and hence
\begin{eqnarray}
	Z_1=1.
\label{ZnextG}
\end{eqnarray}

Now we consider the renormalization group flow in the next-to-leading order.
>From (\ref{ZnextPhi}), we obtain the function
\begin{eqnarray}
	\beta_g = -\varepsilon g \left[1-{g^2 \over 6(4\pi)^3 \varepsilon}(N+2) \right].
\label{betaGnext}
\end{eqnarray}
According to (\ref{betaG}) and (\ref{betaGnext}), we have the differential equation
\begin{eqnarray}
	\mu {\partial g \over \partial \mu}= -\varepsilon g 
	\left[1-{g^2 \over 6(4\pi)^3 \varepsilon}(N+2) \right].
\label{DeffnextG}
\end{eqnarray}
The equation (\ref{DeffnextG}) is solved to give the scale dependence of 
	coupling constant $g$, 
\begin{eqnarray}
	g^2 = \left[{\mu^{2\varepsilon} \over g_0^2}
	+{N+2 \over 6(4\pi)^3 \varepsilon} \right]^{-1},
\label{Gdependnext}
\end{eqnarray}
	where we have chosen $1/g_0^2$ as the integration constant 
	in accordance with (\ref{ZrenG}).
Then (\ref{Gdependnext}) give the renormalization group flow of the coupling constant $g$ 
	in the three-elementary-scalar model in this order.
Substituting (\ref{ZnextChi}) and (\ref{ZnextPhi}) 
	into (\ref{gammaChi}) and (\ref{gammaPhi}), 
	we obtain the anomalous dimensions
\begin{eqnarray}
	\gamma_\chi &=& {g^2 \over 6(4\pi)^3},
\label{gammaChiNext}
\\	\gamma_\phi &=& {Ng^2 \over 6(4\pi)^3},
\label{gammaPhiNext}
\end{eqnarray}
where $g^2$ is given by (\ref{Gdependnext}).

Now let us consider the scale dependence of the masses $m$ and $M$ 
	in the next-to-leading order.
The functions $\beta_m$ and $\beta_M$ are given by
\begin{eqnarray}
	\beta_m &=& -{g^2 m \over 3(4\pi)^3} \left( 1+ {3M^2 \over 2m^2} \right),
\label{betamNext}
\\	\beta_M &=& {Ng^2 M \over 6(4\pi)^3} \left( 1-{6m^2 \over M^2} \right),
\label{betaMNext}
\end{eqnarray}
	from (\ref{Znextm}), (\ref{ZnextM}), (\ref{gammaChiNext}) and (\ref{gammaPhiNext}).
Using (\ref{betamNext}) and (\ref{betaMNext}) for (\ref{betam}) and (\ref{betaM}), 
	we have the differential equations
\begin{eqnarray}
	\mu {\partial m \over \partial \mu} 
	&=& -{g^2 m \over 3(4\pi)^3} \left( 1+ {3M^2 \over 2m^2} \right),
\label{DiffNextm}
\\	\mu {\partial M \over \partial \mu} 
	&=& {Ng^2 M \over 6(4\pi)^3} \left( 1-{6m^2 \over M^2} \right).
\label{DiffNextM}
\end{eqnarray}
Hence we obtain the scale dependence of masses $m$ and $M$, 
\begin{eqnarray}
	m^2 &=& \left\{ 1- {3g^2 \over (4\pi)^3 \varepsilon}
	- {20 \over N}\ln \left[1- {Ng^2 \over 6(4\pi)^3 \varepsilon} \right] \right\}m_0^2
	+{g^2 \over 2(4\pi)^3 \varepsilon}M_0^2,
\label{mnext}
\\	M^2 &=& {Ng^2 \over (4\pi)^3 \varepsilon} 
	\left\{1- {36 \over N}
	+{8 \over N} \left[2-{27(4\pi)^3 \varepsilon \over N g^2} \right]
	\ln \left[1- {Ng^2 \over 6(4\pi)^3 \varepsilon} \right] \right\}m_0^2
\cr&&	+\left\{1+ {(16-N)g^2 \over 6(4\pi)^3 \varepsilon}
	+{16 \over N} \left[1-{Ng^2 \over 6(4\pi)^3 \varepsilon} \right]
	\ln \left[1-{Ng^2 \over 6(4\pi)^3 \varepsilon} \right] \right\}M_0^2,
\label{Mnext}
\end{eqnarray}
	where $g^2$ is given by (\ref{Gdependnext}).
Equations. (\ref{mnext}) and (\ref{Mnext}) give the renormalization group flow 
	for the masses $m$ and $M$ in the next-to-leading order.

\subsection{The composite model in the next-to-leading order}

Here we show that the coupling constant $g$ in the scalar-composite-scalar model 
	persists to be independent of the scale $\mu$ still in the next-to-leading order 
	in $1/N$.
If we use the compositeness condition (\ref{CC}) with (\ref{ZnextPhi}), we obtain
\begin{eqnarray}
	g^2 = {6(4\pi)^3 \varepsilon \over N+2}.
\label{CompGnext}
\end{eqnarray}
We can directly see that the formed $g^2$ in (\ref{CompGnext}) is independent of 
	the scale $\mu$.
Substituting (\ref{CompGnext}) into (\ref{betaGnext}), we find
\begin{eqnarray}
	\beta_g =0,
\label{betaCompNLO}
\end{eqnarray}
	which implies that the scale independence is again a smooth limit of 
	the running in general non-composite case.

On the other hand, the masses $m$ and $M$ are given by
\begin{eqnarray}
	m^2 &\rightarrow& \left(1-{18 \over N}-{20\ln N \over N} \right)m_0^2
	+{3 \over N}M_0^2,
\label{m2ComLim}
\\	M^2 &\rightarrow& \left(6-{216 \over N}-{20\ln N \over N} \right)m_0^2
	+\left({6(4\pi)^3 \varepsilon \mu^{2\varepsilon} \over Ng_0^2}
	+{16 \over N} \right)M_0^2,
\label{M2ComLim}
\end{eqnarray}
	in the compositeness limit (\ref{ComLimt}) in this order.
In this case, however, the expressions
$
	1-{Ng^2/ 6(4\pi)^3 \varepsilon},
$
	inside the logarithms in (\ref{mnext}) and (\ref{Mnext}) vanish, 
	which means it is of $O(1/N)$.
Thus the terms behave like $(\ln N)/N$, in spite that we assigned $O(1/N)$ at first.
The summations over infinite series of the diagrams give rise to the little greater 
	contribution than we expect.
Fortunately this is smaller than the leading order contribution.
The expressions in (\ref{m2ComLim}) and (\ref{M2ComLim}) seem to diverge
	if $M_0\rightarrow\infty$ as is required by (\ref{ComLimt}) for finite $F$,
	where $F$ is the bare coupling constant in (\ref{CompL}).
On the other hand, for $F\rightarrow\infty$, $M_0$ is finite, and 
	we have finite renormalized masses $m$ and $M$. 
The limit $F\rightarrow\infty$ corresponds to the case where 
	the bare mass parameter $F^{-1}$ in (\ref{AuxL}) vanish.

\section{Conclusion}
We have studied the scalar-composite-scalar model in $6-2\varepsilon$ dimensions using 
	equivalence to the renormalizable three-elementary-scalar model with 
	the compositeness condition $Z_3=0$, where 
	$Z_3$ is the renormalization constant 
	of the to-be-composite field $\phi$ in the latter.
Then the physical quantities in the former (scalar-composite-scalar model) are calculated 
	via the latter (three-elementary-scalar model).
The latter consists of complex scalar fields $\chi^1$, $\chi^2$ and $\phi$, 
	while, in the former, $\phi$ is becomes a composite of $\chi^1$ and $\chi^2$, 
	according to the compositeness condition.
We note that the compositeness condition $Z_3=0$ in the latter is the limiting case of 
	the bare coupling constant $g_0 \rightarrow \infty$ and the bare mass 
	$M_0 \rightarrow \infty$, 
	while the physical renormalized quantities $g$ and $M$ are finite, 
	where $g$ and $M$ are the coupling constant and the mass of $\phi$, respectively.
By $1/N$ expansion, we have calculated the renormalization constants 
	$Z_1$, $Z_2$, $Z_3$, $Z_m$ and $Z_M$, 
	in the next-to-leading order in the three-elementary-scalar model.
Then we have derived renormalization group functions $\beta_g$, $\beta_m$ and $\beta_M$.
Solving the renormalization group equations, we have obtained the scale dependence of 
	the effective coupling constant $g$ and 
	the masses of scalar fields $\chi^i$ and $\phi$.

With the purpose of examining the scalar-composite-scalar model, 
	we imposed the compositeness condition $Z_3=0$ in 
	the renormalizable three-elementary-scalar model.
Then, we have the composite scalar field $\phi$ which is comprised of 
	the elementary complex scalars $\chi^1$ and $\chi^2$.
In this composite model, the effective coupling constant $g$ is independent of 
	the scale $\mu$ in the next-to-leading order as was emphasized 
	in the previous letter \cite{ScaleComp}.
The scale independence of coupling constant $g$ is a smooth limit of 
	the running in general non-composite case.
We expect that this scale independence of the coupling constant of the composite field 
	is common to various models to all order.
However, the mass $M$ of the composite field depends on the scale $\mu$, 
	even if we impose the compositeness condition.
If we further impose the condition $Z_M=0$ in addition to 
	the compositeness condition $Z_3=0$, 
	then the mass $M$ also becomes independent of the scale, and 
	we have obtained the relation $M^2 =6m^2$ in the leading order.
It seems that the vanishing conditions for the renormalization constants precisely 
	correspond to the scale independence of the corresponding quantities.

\appendix 
\section{Divergences in the next-to-leading order }

Here we show the calculations of the leading divergence of each diagram
	in the renormalization parts in the next-to-leading order.
The contribution of diagram in Fig.~\ref{f5} is given by
\begin{eqnarray}
	D(q^2) = 
	\left[{Ng^2 \over 6(4\pi)^3 \varepsilon} \right]^n
	\left[1-{\mu^{2\varepsilon} \over (-q^2)^\varepsilon} \right]^n
	\left[{1 \over -q^2}+{6nm^2-(n+1)M^2 \over (-q^2)^2} \right],
\label{D}
\end{eqnarray}
	where $n$ is the number of $\chi^i$-loops on the $\phi$-line, 
	and $q$ is the momentum of $\phi$ on external line.
The contribution from the self-energy part of $\chi^i$ in Fig.~\ref{f3} is given by
\begin{eqnarray}
	S(k^2, \varepsilon)
	&=& g^2 \mu^{2\varepsilon} \sum_{n=0}^\infty 
	\int {d^{3-\varepsilon} q \over i(2\pi)^{3-\varepsilon}}{1 \over m^2-(q+k)^2}D(q^2)
\cr
	&=& {1 \over N}\sum_{n=0}^\infty {1 \over n} \left[{Ng^2 \over 6(4\pi)^3 \varepsilon} \right]^n
	 \left\{ 1- \left[1+{-\mu^{2\varepsilon} \over (-k^2)^\varepsilon} \right]^n \right\}
\cr
&&
	\times \left\{k^2+3 \left[(6n-7)m^2-nM^2 \right] \right\},
\end{eqnarray}
	where $k$ is the momentum of $\chi^i$ on the external line.
Then the divergent part of $S(k^2, \varepsilon)$ is given by
\begin{eqnarray}
	S(k^2, \varepsilon)_{\rm div}
	=
	-{1 \over N}\sum_{n=0}^\infty \left[{Ng^2 \over 6(4\pi)^3 \varepsilon} \right]^n
	\left\{k^2+3 \left[(6n-7)m^2-nM^2 \right] \right\}.
\label{Sdiv}
\end{eqnarray}
The contribution from the self-energy part of $\phi$ in Fig.~\ref{f4} is given by 
\begin{eqnarray}
	\Pi(p^2, \varepsilon)
	&=& 2Ng^2 \mu^{2\varepsilon} \int {d^{3-\varepsilon} k \over i(2\pi)^{3-\varepsilon}}
	{1 \over [m^2-(k+p)^2]}{1 \over (m^2-k^2)^2}
	\left[ S(k^2, \varepsilon)-S(k^2, 0) \right]
\cr
	&=& {2 \over N} \sum_{n=1}^\infty {1 \over n(n+1)}
	\left[{Ng^2 \over 6(4\pi)^3 \varepsilon} \right]^{n+1}
	\left\{ 1- \left[1+ {-\mu^{2\varepsilon} \over (-p^2)^\varepsilon} \right]^{n+1} \right\}
\cr
&&
	\times \left\{ p^2 + 9 \left[6m^2+n(M^2-6m^2) \right] \right\},
\end{eqnarray}
	where $p$ is the momentum of $\phi$ on the external line.
Then the divergent part of $\Pi(p^2, \varepsilon)$ is given by
\begin{eqnarray}
	\Pi(p^2, \varepsilon)_{\rm div}
	= {2 \over N} \sum_{n=1}^\infty {1 \over n(n+1)}
	\left[{Ng^2 \over 6(4\pi)^3 \varepsilon} \right]^{n+1}
	\left\{ p^2 + 9 \left[6m^2+n(M^2-6m^2) \right] \right\}.
\label{Pidiv}
\end{eqnarray}

\begin{figure}
\hskip5cm\epsffile{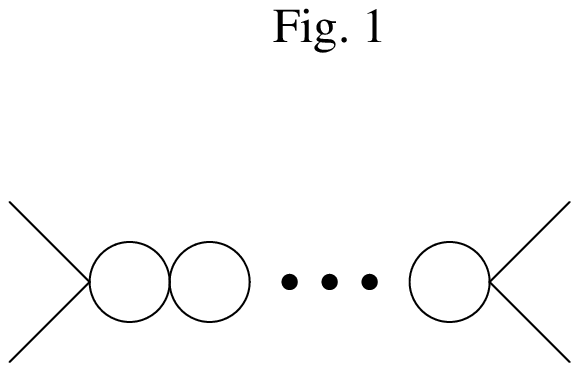}
\caption{ 
The chain diagram. Solid lines indicate propagators of the scalar fields $X^i$.
}
\label{f1}
\end{figure}

\begin{figure}
\hskip5cm \epsffile{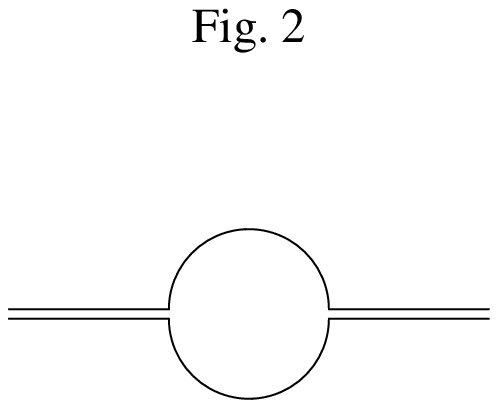}
\caption{
The self-energy diagram of the field $\phi$ at the leading order.
Single and double lines indicate propagators of scalar fields $\chi^i$ and $\phi$,
	respectively.
}
\label{f2}
\end{figure}

\begin{figure}
\hskip5cm \epsffile{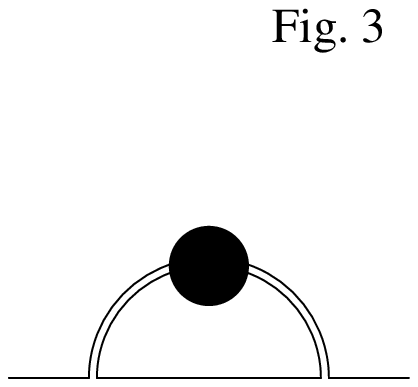}
\caption{
The self energy diagram of $\chi^i$ at the next-to-leading order.
Single and double lines indicate propagators of scalar fields $\chi^i$ and $\phi$,
	respectively.
Blob on the $\phi$-line indicates the insertion of $\chi^i$-loops of arbitrary number as shown in Fig.~5.
}
\label{f3}
\end{figure}

\begin{figure}
\hskip5cm \epsffile{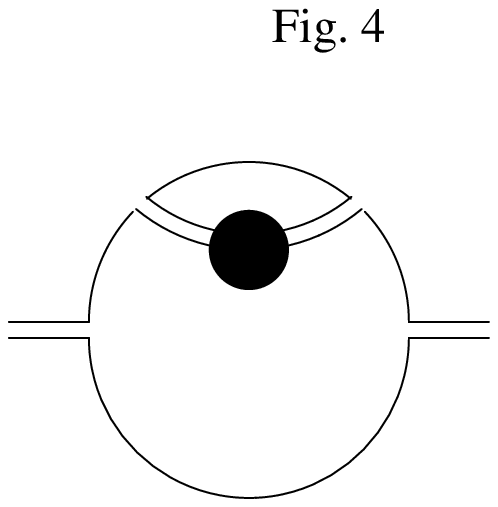}
\caption{
The self energy diagram of $\phi $ at the next-to-leading order.
Single and double lines indicate propagators of scalar fields $\chi^i$ and $\phi$,
	respectively.
Blob on the $\phi$-line indicates the insertion of $\chi^i$-loops of arbitrary number as shown in Fig.~5.
}
\label{f4}
\end{figure}

\begin{figure}
\hskip5cm \epsffile{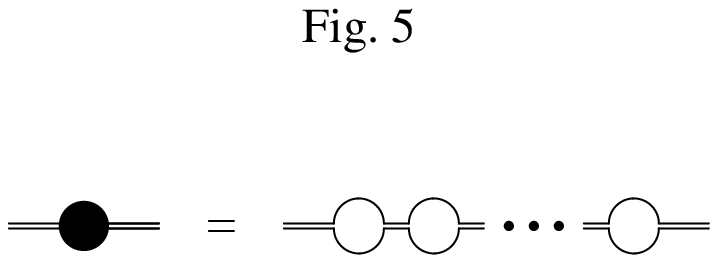}
\caption{
The propagator of $\phi$ inserted with $\chi^i$-loops. 
Single and double lines indicate propagators of scalar fields $\chi^i$ and $\phi$,
	respectively.
Blob on the $\phi$-line indicates the insertion of $\chi^i$-loops of arbitrary number. 
}
\label{f5}
\end{figure}


\begin{thebibliography}{99}

\bibitem{BCS}
J.~Bardeen, L.~N.~Cooper, and J.~R.~Schrieffer, Phys.\ Rev.\ {\bf 106}, 162 (1957);

\bibitem{NJL} 
W.~Heisenberg, Rev.\ Mod.\ Phys.\ {\bf 29}, 269 (1957);
Y.~Nambu and G.\ Jona-Lasinio,	 {Phys.\ Rev.} {\bf 122}, 345 (1961);
For a review, e.g. T.~Hatsuda and T.~Kunihiro,
Phys.\ Rept.\  {\bf 247}, 221 (1994).

\bibitem{ig}
J.~D.~Bjorken, {Ann.\ Phys.} {\bf 24}, 174 (1963);
I.~Bialynicki-Birula, Phys.\ Rev.\ {\bf 130}, 465 (1963);

K.~Akama and T.~Hattori, Phys.\ Lett.\ B {\bf 392}, 383 (1997).


\bibitem{comp}
K.~Akama and H.~Terazawa, INS-Rep-{\bf 257} (1976);
H.~Terazawa, Y.~Chikashige and K.~Akama, Phys.\ Rev.\ D {\bf 15}, 480 (1977);
T.~Saito and K.~Shigemoto,
Prog.\ Theor.\ Phys.\  {\bf 57}, 242 (1977);
L.~Susskind,
Phys.\ Rev.\ D {\bf 20}, 2619 (1979);
O.~W.~Greenberg and J.~Sucher,
Phys.\ Lett.\ B {\bf 99}, 339 (1981);
L.~F.~Abbott and E.~Farhi,
Phys.\ Lett.\ B {\bf 101}, 69 (1981);
K.~Akama and T.~Hattori,
Phys.\ Rev.\ D {\bf 40}, 3688 (1989);
V.A.~Miransky, M.~Tanabashi and K.~Yamawaki,
	 {Phys. Lett.} {\bf B221}, 177 (1989);
	 {Mod.\ {Phys.\ Lett.}} {\bf A4}, 1043 (1989);
W.A. Bardeen, C.T. Hill and M. Lindner, 
	   {Phys. Rev.} D {\bf 41}, 1647 (1990). 


\bibitem{iG}
A.~D.~Sakharov,  Dokl.\ Akad.\ Nauk SSSR {\bf 177}, 70 (1967)
	[{Sov.\ Phys.\ Dokl.} {\bf 12}, 1040 (1968)];
K.~Akama, Y.~Chikashige and T.~Matsuki,
Prog.\ Theor.\ Phys.\  {\bf 59}, 653 (1978);
K.~Akama, Y.~Chikashige, T.~Matsuki and H.~Terazawa,
	 {Prog.\ Theor.\ Phys.} {\bf 60}, 868 (1978);
K.~Akama,  {Prog.\ Theor.\ Phys.} {\bf 60}, 1900 (1978);
A.~Zee, Phys.\ Rev.\ Lett.\ {\bf 42}, 417 (1979);
S.~L.~Adler, Phys.\ Rev.\ Lett.\ {\bf 44}, 1567 (1980). 


\bibitem{BW}
K.~Akama, Lect. Notes in Phys.\ {\bf 176}, 267 (1983);
Prog.\ Theor.\ Phys.\ {\bf 78}, 184 (1987); {\bf 79}, 1299 (1988); 
{\bf 80}, 935 (1988); 
hep-th/0307240 (2003);
V.~A.~Rubakov and M.~E.~Shaposhnikov,
Phys.\ Lett.\ B {\bf 125} (1983) 136;
K.~Akama and T.~Hattori,
Mod.\ Phys.\ Lett.\ A {\bf 15}, 2017 (2000);
G.\ Dvali, G.\ Gabadadze, and M.\ Porrati, Phys.\ Lett.\ {\bf B485}, 208 (2000);
G.~Dvali and G.~Gabadadze,
Phys.\ Rev.\ D {\bf 63}, 065007 (2001);
K.~I.~Maeda, S.~Mizuno and T.~Torii,
Phys.\ Rev.\ D {\bf 68}, 024033 (2003).


\bibitem{CC}
B.~Jouvet, Nuovo Cim. {\bf 5}, 1133 (1956); 
M.~T.~Vaughn, R.~Aaron and R.~D.~Amado, {Phys.\ Rev.} {\bf 124}, 1258 (1961);
A.~Salam, Nuovo Cim. {\bf 25} (1962) 224; 
S.~Weinberg, {Phys.\ Rev.} {\bf 130} (1963) 776;
D.~Luri\'e and A.~J.~Macfarlane,  {Phys.\ Rev.} {\bf 136}, B816 (1964);
T.~Eguchi,  {Phys.\ Rev.} D {\bf 14} (1976) 2755; D {\bf 17}, 611 (1978);
D.~Campbell, F.~Cooper, G.~S.~Guralnik and N.~Snyderman,
Phys.\ Rev.\ D {\bf 19}, 549 (1979);
K.~I.~Shizuya,  {Phys.\ Rev.} D {\bf 21}, 2327 (1980);
R.~W.~Haymaker and F.~Cooper,
Phys.\ Rev.\ D {\bf 19}, 562 (1979);
D.~Luri\'e and G.B.~Tupper,  
Phys.\ Rev.\ D {\bf 47}, 3580 (1993);
K.~Akama,
Phys.\ Rev.\ Lett.\  {\bf 76}, 184 (1996);
K.~Akama, {Phys.\ Lett.} B {\bf 583}, 207 (2004);
A.~Akabane and K.~Akama,
Prog.\ Theor.\ Phys. {\bf 112}, No.\ 4 (2004);
K.~Akama,
Nucl.\ Phys.\ A {\bf 629}, 37C (1998);
K.~Akama and T.~Hattori,
Phys.\ Lett.\ B {\bf 445}, 106 (1998).

\bibitem{history}
D.~J.~Gross and A.~Neveu,
Phys.\ Rev.\ D {\bf 10}, 3235 (1974);
T.~Kugo,
Prog.\ Theor.\ Phys.\  {\bf 55}, 2032 (1976);
K.~Kikkawa,
Prog.\ Theor.\ Phys.\  {\bf 56}, 947 (1976);
J.~Zinn-Justin,
Nucl.\ Phys.\ B {\bf 367}, 105 (1991);
P.~M.~Fishbane, R.~E.~Norton and T.~N.~Truong,
Phys.\ Rev.\ D {\bf 46}, 1768 (1992);
P.~M.~Fishbane and R.~E.~Norton,
Phys.\ Rev.\ D {\bf 48}, 4924 (1993);
J.~A.~Gracey,
Phys.\ Lett.\ B {\bf 308}, 65 (1993);
Phys.\ Lett.\ B {\bf 342}, 297 (1995).

\bibitem{ScaleComp}
K.~Akama and T.~Hattori,
Phys.\ Rev.\ Lett.\  {\bf 93}, 211602-1 (2004).


\end{thebibliography}
\end{document}